\documentclass[journal]{IEEEtran}
\IEEEoverridecommandlockouts

\usepackage{cite}
\usepackage{amsmath,amssymb,amsfonts}
\usepackage{algorithmic}
\usepackage{algorithm}
\usepackage{graphicx}
\usepackage{textcomp}
\usepackage{xcolor}
\usepackage{booktabs}
\usepackage{enumitem}
\usepackage{multirow}
\usepackage{url}
\usepackage{tikz}
\usepackage{pgfplots}
\pgfplotsset{compat=1.18}
\usetikzlibrary{shapes.geometric,arrows,arrows.meta,positioning,fit,backgrounds,patterns}
\usepackage[colorlinks=true,linkcolor=blue,citecolor=blue,urlcolor=blue]{hyperref}

\def\BibTeX{{\rm B\kern-.05em{\sc i\kern-.025em b}\kern-.08em
    T\kern-.1667em\lower.7ex\hbox{E}\kern-.125emX}}

\newcommand{\dSABRE}{\textsc{dSABRE}}

\newcommand{\TeleSABRE}{\textsc{TeleSABRE}}
\newcommand{\SABRE}{\textsc{SABRE}}
\newcommand{\LightSABRE}{\textsc{LightSABRE}}

\newcommand{\arch}{\mathcal{A}}
\newcommand{\dphys}{d_{\mathrm{phys}}}
\newcommand{\dintra}{d_{\mathrm{intra}}}
\newcommand{\dcore}{d_{\mathrm{core}}}
\newcommand{\logi}[1]{\mathrm{q}_{#1}}
\newcommand{\phys}[1]{\mathrm{p}_{#1}}

\begin{document}

\title{dSABRE: A SABRE-Style Router for Multi-Core Distributed Quantum Computers}

\author{Sanjiang~Li~\IEEEauthorrefmark{1}%
\thanks{S. Li is with the Centre for Quantum Software and Information,
  University of Technology Sydney, Australia
  (e-mail: sanjiang.li@uts.edu.au).}%
\thanks{\IEEEauthorrefmark{1}ORCID: 0000-0002-3332-2546.}}

\maketitle

%%─────────────────────────────────────────────────────────────────────────────
\begin{abstract}
Minimising EPR consumption is the dominant objective when routing a
quantum circuit on a distributed quantum computer (DQC).  We
present \dSABRE{}, a \SABRE{}-style router for multi-core processors
that, on each iteration of a lookahead-driven loop, first resolves
any intra-core front-layer gates by SWAP scoring and only
falls back to scoring inter-core teleportation candidates when the
intra-core front is empty.  Three mechanisms drive the
improvement over the state of the art: a five-term
\emph{gate-centric} teleportation score that generalises the local
SWAP heuristic to the inter-core setting, whose explicit
capacity-penalty term keeps the scorer from teleporting into
saturated cores; a proactive congestion-relief pass that redistributes
idle qubits out of high-demand cores before deadlock; and a BFS-layer
construction of the inter-core extended set that respects DAG
dependencies layer by layer rather than mixing wires in topological
order.  Across 18 MQT-Bench circuits
at 25, 36, and 64 logical qubits, \dSABRE{} reduces geometric-mean
EPR consumption by $41$--$44\%$ over \TeleSABRE{} and by $16$--$68\%$
over the gate-teleportation-based \texttt{pytket-dqc}, using
standard Qiskit \texttt{SabreLayout} for the initial layout.  A
large-circuit QFT sweep at 100--360 qubits confirms scalability.  Code and online appendices are available at
\url{https://github.com/ebony72/dsabre}.
\end{abstract}

\begin{IEEEkeywords}
distributed quantum computing, circuit routing, qubit teleportation,
SABRE, EPR minimisation, qubit mapping, lookahead heuristic
\end{IEEEkeywords}

%%─────────────────────────────────────────────────────────────────────────────
\section{Introduction}
\label{sec:intro}

The near-term path to fault-tolerant quantum computing requires processors with thousands of high-fidelity qubits.
No single monolithic chip can economically deliver this scale today.
Instead, the field is converging on \emph{distributed quantum computers}
(DQC)~\cite{cuomo2020towards}: multiple
small quantum processing units (QPUs, also called cores) interconnected by photonic or microwave
links that support remote entanglement generation.

Executing a quantum circuit on a DQC requires a \emph{distributed compilation}
pipeline that, like its monolithic counterpart, can be decomposed into three
problem classes~\cite{escofet2024review}.
\emph{(i) Qubit allocation} statically assigns logical qubits to cores --- e.g.\ via
hypergraph partitioning~\cite{andres2019automated,pytketdqc2024}.
\emph{(ii) Communication-primitive selection} decides how each non-local 2Q gate
is realised: \emph{state teleportation} moves a logical qubit across an
inter-core link~\cite{bennett1993teleporting}; \emph{gate teleportation} (telegate /
EJPP / cat-entanglers) executes a single non-local gate without moving the
qubit~\cite{eisert2000optimal,yimsiriwattana2004generalized}, optionally
amortising a sequence of consecutive non-local gates onto one shared
e-bit~\cite{wu2022autocomm}.  Each primitive consumes one or
more EPR pairs, and EPR generation is slow (microseconds to milliseconds),
noisy, and rate-limited in current
hardware~\cite{stephenson2020high,daiss2021quantum}, making EPR minimisation
the dominant compilation objective.
\emph{(iii) Routing and scheduling} orders intra-core SWAPs, teleports, and
EPR-generation requests in time, subject to communication-port capacity and
finite e-bit rates~\cite{ferrari2021compiler,baker2020time}.

\dSABRE{} addresses the routing component of stage~(iii) --- ordering
intra-core SWAPs and teleports --- and leaves scheduling (EPR-generation
timing, port reservation) to downstream passes; within stage~(ii) it
uses only state teleportation, and it is complementary to the
static-allocation methods of stage~(i).  Within this scope the closest prior work is the family of
\SABRE{}-style~\cite{li2019tackling} distributed routers, which score
teleportation candidates with a heuristic that combines an immediate
front-layer distance reduction and a lookahead term over a fixed window of
upcoming gates~\cite{luo2025dmaps}.
The most recent and best-performing of these is
\TeleSABRE{}~\cite{russo2025telesabre}, which uses a Dijkstra computation on a
contracted communication graph to estimate inter-core routing cost.  We observe
one substantive limitation: congestion is handled defensively.
\TeleSABRE{} discourages landing in saturated cores via a 
full-core penalty and recovers from deadlock via a safety-valve mode
that drops most of the lookahead, but it never proactively
\emph{evicts} idle qubits out of crowded cores ahead of demand.  On
dense or large instances this recovery either burns additional EPR
pairs or fails to converge altogether.

\textbf{Contributions.}

\begin{itemize}

\item \textbf{A five-term gate-centric teleportation score.}
\LightSABRE{}~\cite{zou2024lightsabrelightweightenhancedsabre} scores
a SWAP as the front-layer distance gain $\Delta_F$ plus a
decay-weighted lookahead ${\Delta}_E$, both in $\Delta$-form
(old minus new distance) on gates sharing a qubit with the move.
\dSABRE{} carries the same two $\Delta$-form terms into the
inter-core scorer and adds three teleport-specific terms: a staging
SWAP cost $d_\mathrm{prep}$ to reach a comm port, a
\emph{core-capacity penalty} $c_\mathrm{cap}$ that discourages
landing in nearly-full cores, and a hop-direction reward
$g_\mathrm{hop}$.  Ablation (Section~\ref{sec:ablation}) shows the
capacity term carries the score: removing it inflates geometric-mean
EPR by $+59.9\%$ on the 25-qubit suite and $+58.7\%$ on the 64-qubit
suite, more than any other single mechanism; removing the lookahead
adds another $+28.2\%$ on the 25-qubit suite.

\item \textbf{Proactive congestion relief.}
\dSABRE{} maintains an explicit demand vector over cores and
proactively teleports idle qubits away from high-demand, low-capacity
cores before bottlenecks form, scoring relief moves within the same
teleportation objective.  Disabling relief inflates the 64-qubit geometric-mean EPR by
$+23.4\%$, with the dense AE and QFT circuits more than doubling
(Section~\ref{sec:ablation}); on the 64-qubit Random circuit,
\dSABRE{} completes routing where \TeleSABRE{} fails to converge.
%A checkpoint--rollback wrapper around \LightSABRE{}'s release-valve
%idea~\cite{zou2024lightsabrelightweightenhancedsabre} provides a
%safety net for marginal SabreLayout seeds but does not fire on the
%best-EPR runs.

\item \textbf{BFS-layer inter-core extended set.}
The inter-core scorer builds its lookahead set layer by layer over
the remaining DAG so that gates sharing a wire with the front are
captured before unrelated wires displace them, with each gate's BFS
depth driving the decay exponent $\gamma^{\mathrm{dep}(g)}$.
Substituting a topological-order extended set increases geometric-mean
EPR by $3$--$11\%$ across the three suites, with the gain growing with
circuit size (Section~\ref{sec:ablation}).

\item \textbf{Comprehensive empirical validation.}
Across MQT-Bench suites at 25, 36, and 64 qubits on multi-core grid
architectures, \dSABRE{} reduces geometric-mean EPR consumption by
$41$--$44\%$ over \TeleSABRE{}~\cite{russo2025telesabre} and by
$16$--$68\%$ over \texttt{pytket-dqc}~\cite{pytketdqc2024}, using only
off-the-shelf \texttt{SabreLayout} for initial qubit placement.  Ablation
studies, cost-ratio sweeps, and a scalability sweep up to 360-qubit QFT
confirm that the gains are mainly driven by the heuristic design and congestion
mechanisms.

\end{itemize}

\textbf{Paper organisation.}
Section~\ref{sec:background} fixes notation and recaps the SABRE
heuristic.  Section~\ref{sec:dsabre} presents \dSABRE{}: the routing
loop, the intra-core and inter-core scoring rules, the BFS-layer extended set, 
proactive congestion relief, the checkpoint--rollback escape, and a complexity analysis.  
Section~\ref{sec:experiments}
evaluates \dSABRE{} on the 25q, 36q, and 64q MQT-Bench suites
against \TeleSABRE{} and \texttt{pytket-dqc}, ablates the design
choices, characterises the initial-layout pipeline, reports compile
time, and runs a 100--360q QFT scalability sweep.
Section~\ref{sec:related} positions \dSABRE{} against related work;
Section~\ref{sec:future} outlines directions for further work;
Section~\ref{sec:conclusion} concludes.

%%─────────────────────────────────────────────────────────────────────────────
\section{Background}
\label{sec:background}

\subsection{Quantum Circuits}
\label{sec:qcircuits}

A quantum circuit acts on $n$ \emph{qubits} through a sequence of \emph{gates}.
\emph{One-qubit (1Q) gates} (Hadamard, rotations, \ldots) act on a single qubit.
\emph{Two-qubit (2Q) gates} entangle pairs: the \emph{CX} (CNOT) gate is the
standard primitive and the building block for most multi-qubit operations.
Together, CX and arbitrary 1Q gates form a universal gate set, so any
$n$-qubit unitary can be compiled into a circuit over this basis.
The \emph{SWAP} gate exchanges the states of two adjacent qubits; it decomposes
into three CX gates and is used by routing algorithms to migrate logical qubit
states along the coupling graph.

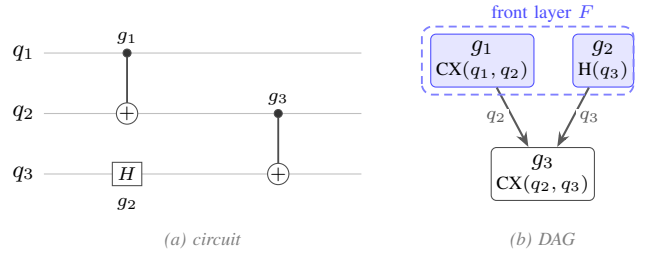
\begin{figure}[t]
  \centering
  % DAG construction example (Sec. II-A)
% Circuit: g1=CX(q1,q2),  g2=H(q3),  g3=CX(q2,q3)
% Edges: g1->g3 (qubit q2),  g2->g3 (qubit q3)
% Shows: parallel front-layer gates g1,g2 and their shared successor g3.
\begin{tikzpicture}[
  every node/.style={font=\small},
  gate/.style={rectangle, rounded corners=2pt, draw=black!70, fill=white,
               minimum width=6mm, minimum height=7mm, inner sep=2pt,
               align=center},
  fgate/.style={gate, draw=blue!70, fill=blue!10},   % front-layer gates
  dep/.style={-{Stealth[length=2mm,width=1.5mm]}, thick, black!65},
  lbl/.style={font=\scriptsize, fill=white, inner sep=1pt},
]

%% ── left panel: circuit as wire diagram ──────────────────────────────
\begin{scope}[xshift=-1cm]
  % qubit wires
  \foreach \y/\q in {1.6/q_1, 0.8/q_2, 0/q_3} {
    \draw[gray!50] (-0.1,\y) -- (4.1,\y);
    \node[left, font=\small] at (-0.1,\y) {$\q$};
  }
  % g1 = CX(q1,q2) at x=1.0
  \draw[black!70, line width=0.7pt] (1.0,1.6) -- (1.0,0.8);
  \node[circle, draw=black!70, fill=black!80, minimum size=3pt,
        inner sep=0pt] at (1.0,1.6) {};          % control
  \node[circle, draw=black!70, fill=white, minimum size=8pt,
        inner sep=0pt, font=\scriptsize] at (1.0,0.8) {$+$};   % target
  \node[above, font=\scriptsize] at (1.0,1.6) {$g_1$};

  % g2 = H(q3) at x=1.0
  \node[rectangle, draw=black!70, fill=white, minimum size=7pt,
        inner sep=2pt, font=\scriptsize] at (1.0,0) {$H$};
  \node[above, font=\scriptsize] at (1.0,-0.6) {$g_2$};

  % g3 = CX(q2,q3) at x=3.0
  \draw[black!70, line width=0.7pt] (3.0,0.8) -- (3.0,0);
  \node[circle, draw=black!70, fill=black!80, minimum size=3pt,
        inner sep=0pt] at (3.0,0.8) {};
  \node[circle, draw=black!70, fill=white, minimum size=8pt,
        inner sep=0pt, font=\scriptsize] at (3.0,0) {$+$};
  \node[above, font=\scriptsize] at (3.0,0.8) {$g_3$};

  \node[below, font=\scriptsize\itshape, text=gray] at (2.0,-0.65) {(a) circuit};
\end{scope}

%% ── right panel: DAG ─────────────────────────────────────────────────
\begin{scope}[xshift=5.5cm, yshift=0.9cm]
  % nodes
  \node[fgate] (g1) at (-0.8, 0.6) {$g_1$\\[-1pt]\scriptsize CX$(q_1,q_2)$};
  \node[fgate] (g2) at ( 0.8, 0.6) {$g_2$\\[-1pt]\scriptsize H$(q_3)$};
  \node[gate]  (g3) at ( 0.0,-0.9) {$g_3$\\[-1pt]\scriptsize CX$(q_2,q_3)$};

  % dependency edges
  \draw[dep] (g1) -- node[lbl, left=1pt] {$q_2$} (g3);
  \draw[dep] (g2) -- node[lbl, right=1pt]{$q_3$} (g3);

  % front-layer annotation
  \draw[blue!50, rounded corners=4pt, densely dashed, line width=0.7pt]
      (-1.6,0.15) rectangle (1.2,1.05);
  \node[blue!70, font=\scriptsize] at (0, 1.22) {front layer $F$};

  \node[below, font=\scriptsize\itshape, text=gray] at (0,-1.55) {(b) DAG};
\end{scope}

\end{tikzpicture}
  \caption{DAG for a three-gate circuit on qubits $q_1,q_2,q_3$.
    $\mathrm{CX}(q_1,q_2)$ and $\mathrm{H}(q_3)$ share no qubit and have
    no incoming edges, so both appear in the initial front layer $F=\{g_1,g_2\}$
    and may execute in parallel.
    $\mathrm{CX}(q_2,q_3)$ depends on both via qubits $q_2$ and $q_3$;
    it enters $F$ only after $g_1$ and $g_2$ have been executed.
    Circuit depth is~2 (two sequential layers).}
  \label{fig:dag-example}
\end{figure}

A circuit is represented as a \emph{directed acyclic graph} (DAG) in which each
node is a gate and each directed edge records a data dependency (qubit produced
by gate $g_1$ consumed by gate $g_2$).
The construction rule is: for every qubit $q$, connect each gate acting on $q$
to the \emph{next} gate acting on $q$ by a directed edge; the edge means the
second gate cannot begin until the first has written its output on $q$.
One-qubit gates therefore contribute at most one outgoing edge per operand,
while 2Q gates contribute one per operand pair.
Figure~\ref{fig:dag-example} shows a small example, where \emph{circuit depth} is the length of the longest path through the DAG.

Physical qubits are arranged in a \emph{coupling graph} $G_\mathrm{phys}=(P,E_\mathrm{phys})$;
a 2Q gate can execute only if its two operands occupy adjacent vertices.
A circuit is written in terms of \emph{logical qubits} $\logi{1},\ldots,\logi{n}$;
a \emph{layout} $\ell:\logi{i}\!\mapsto\!\phys{p}$ assigns each to a physical
slot.
\emph{Qubit routing} inserts SWAP gates so that the two operands of every 2Q
gate reach adjacent physical qubits before it executes.

\subsection{Quantum Teleportation}
\label{sec:teleportation}

\emph{Quantum teleportation}~\cite{bennett1993teleporting} transfers the
state of a qubit from a sender to a receiver---in the DQC setting, from
one core to another over an inter-core link---using a pre-shared
\emph{EPR pair} (a maximally entangled two-qubit state) and two
classical bits.  The source qubit's state is destroyed at the sender
and reconstructed at the receiver; no physical qubit travels across
the link.  Each EPR pair is consumed upon use, so the \emph{EPR count}
is the primary inter-core routing cost in distributed quantum
compilation.  Throughout this paper, ``EPR'' as a unit (e.g.\ ``$10$
EPRs'') denotes EPR-pair count; ``EPR pair'' is used when emphasising
the physical entangled state itself.

Two inter-core teleportation variants arise in DQC compilation.
\emph{Teledata} moves a logical qubit to a new core (one EPR pair consumed),
after which the qubit participates in local gates there.
\emph{Telegate}~\cite{eisert2000optimal} executes a non-local 2Q gate directly
across two cores via a shared EPR pair, without physically relocating either
qubit.
\dSABRE{} uses teledata exclusively; \TeleSABRE{} exploits both variants.

\subsection{Distributed Quantum Architecture}
\label{sec:arch}

\begin{figure*}[t]
  \centering
  \resizebox{0.6\linewidth}{!}{\input{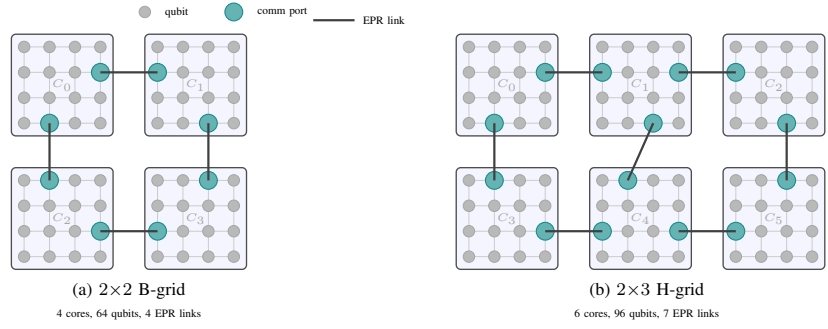}}
  \caption{DQC architectures used in this paper (B-grid and H-grid
    families introduced by \TeleSABRE{}~\cite{russo2025telesabre}).
    Small grey circles are qubits; teal circles are inter-core communication
    ports; bold lines are EPR-channel links between cores.
    Each core is a $4{\times}4$ grid.
    (a)~$2{\times}2$ B-grid.
    (b)~$2{\times}3$ H-grid.}
  \label{fig:arch}
\end{figure*}

We model a DQC architecture $\arch$ abstractly as a weighted graph
$G_\arch=(V,E,w)$ whose vertices are physical qubits.
The vertex set is partitioned into $K$ \emph{cores} $V = V_1\sqcup\cdots\sqcup V_K$;
intra-core edges $E_{\mathrm{intra}}$ describe in-core connectivity and carry
weight~$1$, and inter-core edges $E_{\mathrm{inter}}$ connect designated
\emph{communication ports} on neighbouring cores and carry a much larger
weight $w_{\mathrm{link}} \gg 1$ reflecting the cost of an EPR pair.
A communication port is a physical qubit like any other and may also
hold a logical (data) qubit between teleportations; the
``communication'' label only indicates that the port is one end of an
inter-core EPR link, not that the slot is reserved.
\dSABRE{} treats $G_\arch$ as a black box and only queries it through three
distance tables: physical $\dphys$, intra-core $\dintra$, and core-graph
$\dcore$.
The router therefore extends to \emph{any} DQC topology --- arbitrary
intra-core graphs, mixed core sizes, sparse inter-core meshes, hierarchical
trees of clusters, etc. --- by simply supplying these tables.

For experimental simplicity our reference implementation instantiates
$G_\arch$ as a \emph{grid-of-grids}~\cite{russo2025telesabre}: each core
is an $m\!\times\!m$ 2D nearest-neighbour grid, and the cores themselves
form an $r\!\times\!s$ super-grid joined by a small set of inter-core
links between specific boundary qubits.  
Figure~\ref{fig:arch} shows two
concrete topologies used in our experiments.

\paragraph{Hierarchical distance precomputation.}
Because intra-core and inter-core costs decouple, the all-pairs distance
$\dphys$ over $G_\arch$ admits a two-level Dijkstra computation that is
substantially cheaper than running Dijkstra on the full $|V|$-vertex graph.
First, all-pairs distances are computed once \emph{inside} a single core
$V_i$ (size $|V_i|=M$) in $O(M^2 \log M)$ time, giving $\dintra$.
Second, the \emph{core graph} (a multigraph of $K$ super-nodes connected
by inter-core links) is solved in $O(K^2 \log K)$ time, giving $\dcore$.
Third, $\dphys(p,q)$ for any $p\in V_i$, $q\in V_j$ is recovered as
$\dintra(p,\pi_i^*) + \dcore(i,j) + \dintra(\pi_j^*,q)$ minimised over
boundary-port pairs $(\pi_i^*, \pi_j^*)$ realising the optimal core-graph
path.
This brings precomputation from $O((KM)^2 \log(KM))$ down to
$O(KM^2\log M + K^2\log K)$ and is the only architecture-dependent setup
\dSABRE{} performs; routing itself uses $O(1)$ table lookups thereafter.

\subsection{Qubit Allocation and Routing}

Given circuit DAG $C$ and initial layout $\ell:\logi{i}\!\mapsto\!\phys{\sigma(i)}$ ($\sigma$ is a permutation on ${0,1,...,n-1}$ with $n$ the number of physical qubits),
routing inserts the minimum number of SWAP and teleportation operations so every
gate can execute on adjacent physical qubits.
Cost model: each local SWAP costs $c_{\mathrm{swap}}$; each inter-core
teleport costs $c_{\mathrm{tele}}$ (one EPR pair).

\subsection{SABRE}

\LightSABRE{}~\cite{zou2024lightsabrelightweightenhancedsabre} optimises \SABRE{}~\cite{li2019tackling} by selecting the SWAP minimising:
\begin{equation}
  H = \tfrac{1}{|F|}\!\sum_{g\in F}\!\Delta_F(g)
    + w\,\tfrac{1}{|E|}\!\sum_{g\in E}\!\Delta_E(g),
  \label{eq:sabre}
\end{equation}
where $F$ is the front layer, $E$ the extended lookahead set, and
$\Delta(g) = d_{\mathrm{old}}^g - d_{\mathrm{new}}^g$ the reduction in
physical distance between the two qubits of gate $g$ caused by the
candidate SWAP ($d^g$ is shortest-path distance on the coupling graph).
All extended-set gates contribute with equal weight; the $\tfrac{1}{|E|}$
factor normalises the lookahead term to the same scale as the front term.
\SABRE{} also couples its router with an initial-layout optimiser: starting
from a random qubit assignment it routes the circuit $C$ forward, then
immediately routes the reverse circuit $C^{-1}$ using the final layout of
the forward pass as the new starting point; the layout produced by the
reverse pass is fed back as the initial layout for the next forward pass.
After several bidirectional sweeps the heuristic has annealed towards a
layout that is self-consistent with the circuit's gate structure.

%%─────────────────────────────────────────────────────────────────────────────
\section{dSABRE}
\label{sec:dsabre}

\dSABRE{} is a SABRE-style routing algorithm for multi-core
processors.  At the intra-core level its SWAP-selection objective
inherits the front-layer / extended-set scoring of
Eq.~\ref{eq:sabre}, with two refinements: a decay-weighted lookahead
($\gamma^{\mathrm{dep}(g)}$ down-weights deep successors) and taint
propagation that excludes qubits already entangled with cross-core
gates (Section~\ref{sec:intra}).  Inter-core movement is handled by
a separate teleportation scorer (Section~\ref{sec:tele}).

Routing assumes a fixed initial layout; we use Qiskit
\texttt{SabreLayout}, and defer the discussion of layout preparation
and the routing pass schedule to Section~\ref{sec:initial_mapping}.

The core invariants and notation are: (i) the layout
$\ell:\logi{q}\!\to\!\phys{p}$ assigns logical to physical qubits and
the inverse $\ell^{-1}$ records which logical qubit (if any) currently
occupies each physical slot; (ii) the working DAG~$W$ holds the
remaining gates (initially the full circuit) and shrinks as gates are
executed; (iii) the front layer~$F$ is the set of gates in $W$ with no
unexecuted predecessors, and the extended set~$E$ is a fixed-size
sliding lookahead beyond $F$.

\subsection{Workflow}
\label{sec:workflow}

\begin{figure*}[t]
\centering
\begin{tikzpicture}[
  node distance=4.5mm and 6mm,
  every node/.style={font=\footnotesize},
  block/.style={draw, rounded corners=1.5pt, align=center,
                minimum height=8mm, minimum width=8mm,
                fill=blue!4, inner sep=2pt},
  decision/.style={draw, diamond, aspect=2.4, align=center,
                   fill=orange!10, inner sep=1pt},
  formula/.style={draw, dashed, rounded corners=1.5pt, align=left,
                  fill=gray!5, inner sep=3pt, font=\scriptsize},
  arrow/.style={-{Stealth[length=2mm]}, thick},
]
% top row: input → drain → classify
\node[block] (input) {Working DAG $W$\\layout $\ell$};
\node[block, below=4mm of input] (drain)
   {\textbf{P1.}~Drain front\\(execute 1Q gates,\\adjacent intra-2Q gates)};
% merge
\node[decision, below=8mm of drain] (prog) {\textbf{P4.}~$|W|$ shrank?};
\node[decision, below=4mm of prog] (comp) {$F$\\empty?};
\node[decision, left=8mm of comp] (cls) {\textbf{P2.}~$F_\mathrm{intra}$\\empty?};
\node[block, right=3mm of comp, fill=green!8] (done) {Done};
% intra branch (top)
\node[block, above=8mm of cls] (intra)
   {\textbf{P3a.}~Best intra-core SWAP\\$\arg\min H_\mathrm{intra}$ (Eq.\,\ref{eq:intra})};
% inter branch (bottom)
\node[block, left=4mm of intra] (inter)
   {\textbf{P3b.}~Best teleport\\$\arg\min s$ (Eq.\,\ref{eq:dsabre})\\+\,relief candidates};

\node[decision, right =8 mm of prog] (deadlock)
   {$L_\mathrm{deadlock}$\\ reached?};
\node[block, above=3 mm of deadlock] (recovery)
   {Restore checkpoint;\\force-hop stuck gate};

% loop back

% arrows
\draw[arrow] (input)  -- (drain);
\draw[arrow] (drain)  -- (prog);
\draw[arrow] (prog) -- (comp) node[midway,right,font=\scriptsize] {yes};
\draw[arrow] (comp) -- (cls) node[midway,right,yshift=2mm,font=\scriptsize] {no};
\draw[arrow] (comp) -- (done) node[midway,above,xshift=-1mm,font=\scriptsize] {yes};
\draw[arrow] (cls) -- (intra)node[midway,left,font=\scriptsize] {no};

\draw[arrow] (cls.west)  -| (inter.south) node[midway,left,font=\scriptsize] {yes};

\draw[arrow] (inter.north)  |- (drain.west);
\draw[arrow] (intra.north) |- (drain.west);

\draw[arrow] (prog) -- (deadlock) node[midway,above,font=\scriptsize] {no};
\draw[arrow] (recovery.north) |- (drain);
\draw[arrow] (deadlock) -- ++ (0mm,-24mm) -| (cls.south)node[midway,left,font=\scriptsize] {no};
\draw[arrow] (deadlock) -- (recovery)node[midway,left,font=\scriptsize] {yes};

% formula callouts
\end{tikzpicture}
\caption{The \dSABRE{} routing workflow.  Each iteration drains the
front layer (P1), classifies the remaining front into intra-core and
inter-core gates (P2), applies one SWAP or one teleport (P3), and
checkpoints when forward progress is made (P4).}
\label{fig:workflow}
\end{figure*}
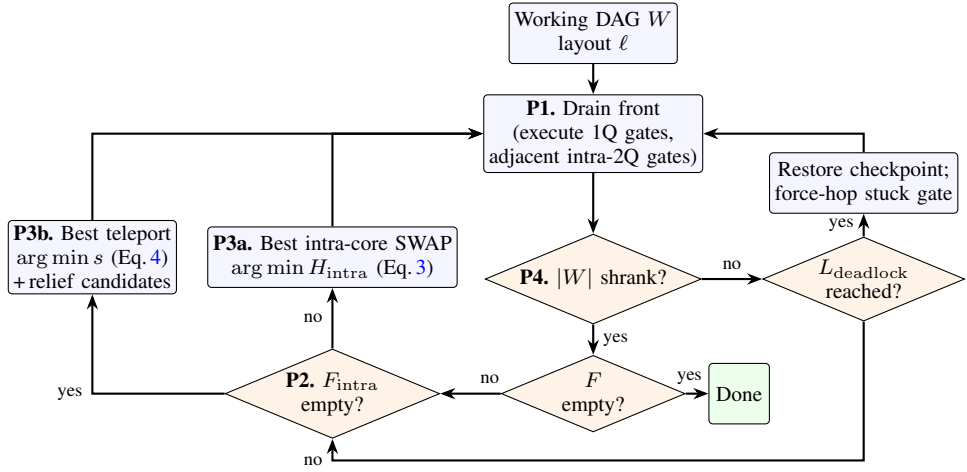

Algorithm~\ref{alg:dsabre} and Figure~\ref{fig:workflow} sketch the
main routing loop.  Each
iteration runs four phases: \textbf{(P1)}~drain $F$ (execute 1Q gates
and any 2Q gate whose operands are already adjacent in the same core,
iterated to a fixed point); \textbf{(P2)}~partition the remaining
front into $F_\mathrm{intra}$ (both operands in one core) and
$F_\mathrm{inter}$ (operands in different cores); \textbf{(P3)}~if
$F_\mathrm{intra}\!\neq\!\emptyset$ apply one intra-core SWAP
(Section~\ref{sec:intra}), else score teleportation candidates over
$F_\mathrm{inter}$ together with proactive relief candidates and apply
the cheapest (Section~\ref{sec:tele});
\textbf{(P4)}~checkpoint $(\ell, W)$ if $|W|$ decreased, otherwise
trigger checkpoint--rollback after $L_\mathrm{deadlock}$ stalled
iterations and abort after $N_\mathrm{backup}^\mathrm{max}$ failed
recoveries.  The two scoring functions operate at different scopes:
the intra-core scorer is per-core, restricted to intra-core
continuations (qubits involved in inter-core gates are tainted), and
front-/extended-set averaged in the \SABRE{} style;
the inter-core scorer evaluates moves across cores using the global
$F_\mathrm{inter}$ and a global BFS-layer extended set
(Section~\ref{sec:bfsext}), with $\tilde{\Delta}_E$ entering as an
un-averaged decay-weighted sum because the cross-core lookahead window
is small and its size already enters implicitly via the
$|E|_{\mathrm{max}}$ cap.

\begin{algorithm}[t]
\caption{\dSABRE{} \texttt{route}$(C, \ell_0)$}
\label{alg:dsabre}
\begin{algorithmic}[1]
\STATE $W \leftarrow$ working copy of circuit DAG $C$;\ \ $\ell \leftarrow \ell_0$
\STATE $C_\mathrm{out} \leftarrow$ empty routed circuit on the physical qubits
\STATE save checkpoint $(\ell, W)$
\WHILE{$W$ has unexecuted gates}
  \STATE drain $F$ (execute 1Q gates and adjacent intra-core 2Q gates; append them to $C_\mathrm{out}$) \hfill\textit{(P1)}
  \IF{$W$ empty} \STATE \textbf{break} \ENDIF
  \STATE compute $F$, partition into $F_\mathrm{intra}, F_\mathrm{inter}$ \hfill\textit{(P2)}
  \IF{$F_\mathrm{intra}\!\neq\!\emptyset$}
    \STATE compute per-core intra-only extended set $E_c$ for each active core $c$ \hfill\textit{(P3a)}
    \STATE pick best SWAP $(u,v)$ within a core (Eq.~\ref{eq:intra})
    \STATE apply SWAP$(u,v)$ to $\ell$;\ \ append SWAP$(u,v)$ to $C_\mathrm{out}$
  \ELSIF{$F_\mathrm{inter}\!\neq\!\emptyset$}
    \STATE compute global $E$ if not cached \hfill\textit{(P3b)}
    \STATE generate candidates over $F_\mathrm{inter}$ and relief moves
    \STATE pick lowest-score candidate (Eq.~\ref{eq:dsabre})
    \STATE apply teleport to $\ell$;\ \ append teleport (staging SWAPs $+$ teledata) to $C_\mathrm{out}$
  \ENDIF
  \IF{$|W|$ decreased}
    \STATE save checkpoint \hfill\textit{(P4)}
  \ELSIF{$L_\mathrm{deadlock}{=}50$ stalled iterations}
    \STATE restore checkpoint;\ \texttt{backup\_plan}$(W,\ell,C_\mathrm{out})$ \hfill\textit{// abort after $N_\mathrm{backup}^\mathrm{max}{=}50$}
  \ENDIF
\ENDWHILE
\RETURN $(\ell, C_\mathrm{out}, \mathrm{metrics})$
\end{algorithmic}
\end{algorithm}

\subsection{Intra-Core SWAP Heuristic}
\label{sec:intra}

When the front contains a gate whose operands share a core but are
not adjacent, \dSABRE{} chooses an intra-core SWAP using the standard
\SABRE{} heuristic, restricted to the relevant core.  Let
$F_\mathrm{c} = \{g\in F_\mathrm{intra} : \mathrm{core}(g)=c\}$ be the
front gates in core~$c$ and $E_\mathrm{c} = (g_0, g_1, \ldots)$ the
corresponding intra-core extended set, an \emph{ordered list} of at most
$L$ gates (default $L{=}20$) collected in topological order.
The traversal stops including a gate the moment either of its qubits
has previously appeared in a cross-core gate in the working DAG
(taint propagation), ensuring $E_\mathrm{c}$ contains only gates
that remain local to core~$c$.
Each gate~$g_i$ is assigned a depth $\mathrm{dep}(g_i)$, the number of
intra-core gates on the longest qubit-wire path from the front layer to
$g_i$ (front layer = depth~0; first successor = depth~1).
The lookahead contribution is
\begin{equation}\label{Delta_E_Tilde}
\tilde{\Delta}_E^\mathrm{c} = \sum_{i=0}^{|E_\mathrm{c}|-1}
\gamma^{\mathrm{dep}(g_i)}(d_{\mathrm{old}}^{g_i} - d_{\mathrm{new}}^{g_i}),
\end{equation}
where $\gamma\in (0,1]$ is the lookahead decay and the distance terms use
the intra-core graph.
The tilde marks the departure from \SABRE{}'s flat extended-set weighting:
rather than treating all lookahead gates equally, the exponential factor
$\gamma^{\mathrm{dep}(g_i)}$ down-weights gates far from the front so that
near-term successors dominate the lookahead signal --- a deeper gate is
more likely to be displaced by intervening routing decisions and so
contributes less reliable information about the right SWAP now.
The same weighted scheme is reused by the inter-core scorer
(Section~\ref{sec:tele}); Section~\ref{sec:bfsext} explains how
$\gamma^{\mathrm{dep}(g)}$ interacts with the BFS-layer extended set,
where $\mathrm{dep}(g)$ is the BFS depth rather than an iteration
index.
For every neighbour pair $(u,v)$ within a core, the score
\begin{equation}
  H_\mathrm{intra} = \frac{\Delta_F^\mathrm{c}}{|F_\mathrm{c}|}
        + w_e\,\frac{\tilde{\Delta}_E^\mathrm{c}}{|E_\mathrm{c}|}
  \label{eq:intra}
\end{equation}
is the SABRE objective for intra-core scoring with the
lookahead-decay weighting of Eq.~\ref{Delta_E_Tilde}.
The lowest-score SWAP is applied; tie-breaking is by enumeration
order.  Because the scorer is restricted to a single core and the
candidate set is the $O(M)$ intra-core neighbour pairs, the
per-core selection is trivially parallel across
active cores.

\subsection{Inter-Core Teleportation Scoring}
\label{sec:tele}

For each gate $g\in F_\mathrm{inter}$ on logical qubits $q_1, q_2$,
let $c_\mathrm{src}$ and $c_\mathrm{tgt}$ be the cores currently
hosting $q_1$ and $q_2$ respectively (i.e.\ $\ell(q_1)\in c_\mathrm{src}$
and $\ell(q_2)\in c_\mathrm{tgt}$, where $\ell$ is the current layout).
Both endpoints are tried as teleportation sources.
For the qubit chosen as source (say $q_1$ at physical position $p_1 = \ell(q_1)$),
the router enumerates every neighbouring core
$c_\mathrm{next}$ and every inter-core link
$(\pi_s, \pi_d)$ between $c_\mathrm{src}$ and $c_\mathrm{next}$.
Each such triple $(q_1, c_\mathrm{next}, (\pi_s,\pi_d))$ is one
\emph{candidate}; Figure~\ref{fig:tele-candidates} illustrates the structure.
Executing a candidate involves two steps: (1)~intra-core SWAPs that move
$q_1$ from $p_1$ to the \emph{staging slot} $n_s$, the neighbour of $\pi_s$
closest to $p_1$; and (2)~a teleportation that consumes the EPR pair on link
$(\pi_s,\pi_d)$ and delivers $q_1$ from $n_s$ to $\pi_d$ in $c_\mathrm{next}$.
Each candidate is scored as:

\begin{figure}[t]
  \centering
  \resizebox{\linewidth}{!}{\input{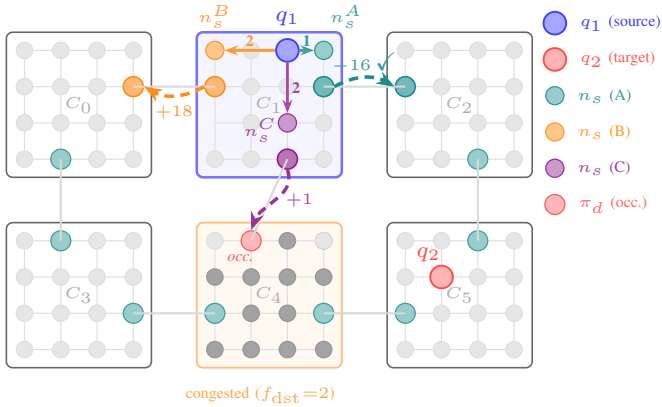}}
  \caption{Running example for inter-core teleportation scoring (2$\times$3
    H-grid).  Candidates \textbf{A}, \textbf{B}, and \textbf{C} below
    correspond row-by-row to Table~\ref{tab:example}.
    Gate $g=(q_1,q_2)$ has $q_1$
    in source core $C_1$ (blue border) and $q_2$ in target core $C_5$.  The
    router considers three candidates, each a triple
    $(q_1,\,c_\mathrm{next},\,(\pi_s,\pi_d))$: \textbf{A}~(teal) routes
    $q_1$ towards $C_2$; \textbf{B}~(orange) routes towards $C_0$, away from
    $C_5$; \textbf{C}~(violet) routes towards congested landing core $C_4$
    (orange border, $f_\mathrm{dst}{=}2$, where $f_\mathrm{dst}$ counts
    free slots in $c_\mathrm{next}$, not in $c_\mathrm{tgt}$).
    Step~(1): intra-core SWAPs move $q_1$ from $p_1$ to staging slot $n_s$
    (count shown on arrow).  Step~(2): a teleportation along link
    $(\pi_s,\pi_d)$ delivers $q_1$ to $\pi_d$ in $c_\mathrm{next}$
    (dashed arc; score $s$ annotated).  Candidate~A wins ($s_A{=}{-16}$) by
    combining the fewest staging SWAPs with positive $\Delta_F$ and hop
    gain and no capacity penalty.  The red port ($\pi_d^C$, labelled
    \textit{occ.}) indicates an occupied landing port, contributing an
    extra eviction SWAP to $d_\mathrm{prep}^C$.}
  \label{fig:tele-candidates}
\end{figure}
\begin{equation}
  s = \underbrace{d_{\mathrm{prep}}}_{\text{staging}} +
      \underbrace{c_{\mathrm{cap}}}_{\text{capacity}} -
      \underbrace{g_{\mathrm{hop}}}_{\text{hop gain}} -
      \underbrace{\Delta_F}_{\text{front gain}} -
      \underbrace{w_e\,\tilde{\Delta}_E}_{\text{lookahead}},
  \label{eq:dsabre}
\end{equation}
where lower scores are preferred.  Each term is defined as follows.

\textbf{Staging cost} $d_{\mathrm{prep}}$ is the number of intra-core SWAPs
needed to move $q_1$ from $p_1$ to staging slot $n_s$, plus eviction SWAPs
to free the comm ports if occupied:
\[
  d_{\mathrm{prep}} = d_{\mathrm{intra}}(p_1,\,n_s)
                    + d_{\mathrm{evict}}(\pi_s)
                    + d_{\mathrm{evict}}(\pi_d).
\]

\textbf{Capacity penalty} $c_{\mathrm{cap}} = c_{\mathrm{pen}}\,\max(0,\,\tau - f_{\mathrm{dst}})$
discourages routing into nearly-full cores.
$f_{\mathrm{dst}}$ is the number of free slots in the immediate landing
core $c_\mathrm{next}$ (not the final partner-qubit core $c_\mathrm{tgt}$),
$\tau$ the capacity threshold, and $c_{\mathrm{pen}}$ a cost multiplier.
This prevents progressive crowding of hot cores that leads to deadlock.

\textbf{Front-layer gain} $\Delta_F = d_{\mathrm{old}}^{g} - d_{\mathrm{new}}^{g}$
is the reduction in weighted physical distance between $q_1$ and $q_2$
after the teleport, computed on the full-chip graph $G_\arch$ (intra-core
edges weight~1, inter-core links weight~$w_{\mathrm{link}}$).
There is exactly one front-layer gate involving the moving qubit $q_1$:
because any two gates that share a qubit are ordered by a dependency
edge in the DAG, at most one of them can be unblocked (i.e.\ in $F$)
at any time, and $g = (q_1,q_2) \in F_{\mathrm{inter}}$ is already
that gate by the premise of this scoring step.

\textbf{Hop gain} $g_{\mathrm{hop}} = w_h\,(\dcore(c_{\mathrm{src}},c_{\mathrm{tgt}}) - \dcore(c_{\mathrm{next}},c_{\mathrm{tgt}}))$
is a position-independent directional reward that supplements
$\Delta_F$ when the landing port $\pi_d$ is far from the best staging
position for the next hop, causing $\Delta_F$ to under-report
core-level progress.  The term is largely redundant with $\Delta_F$
on small topologies (the B-grid suites tie exactly with
$g_{\mathrm{hop}}$ ablated) and contributes only on larger H-grids
($+1.9\%$ geometric-mean EPR on the 64-qubit suite,
Section~\ref{sec:ablation}); its importance is expected to grow
on architectures with larger cores, where the landing port can sit
many SWAPs away from the next staging position.

\textbf{Decay-weighted lookahead.}
$\tilde{\Delta}_E$ is the same decay-weighted lookahead introduced
in Section~\ref{sec:intra} (Eq.~\ref{Delta_E_Tilde}), reused here
with two adaptations for the inter-core scorer: $E$ is now the
inter-core extended set (Section~\ref{sec:bfsext}) rather than the
per-core taint-propagated list, and the sum is restricted to gates
involving the teleported qubit, since only those gates are affected
by the candidate move.  The same $\gamma$ controls how rapidly
deeper successors are discounted.

\paragraph{Running example.}
Figure~\ref{fig:tele-candidates} illustrates Eq.~\ref{eq:dsabre} on
the $2{\times}3$ H-grid with defaults $w_h{=}5$, $w_{\mathrm{link}}{=}10$,
$\tau{=}3$, $c_{\mathrm{pen}}{=}15$.  Gate $g{=}(q_1,q_2)$ has
$\dcore(C_1,C_5){=}2$ and is the only pending gate on $q_1$, so
$\Delta_F$ is evaluated from $g$ alone and $\tilde{\Delta}_E{=}0$
(empty extended set).  With $p_1{=}(0,2)\in C_1$ and
$p_2{=}(1,1)\in C_5$, the shortest path on $G_\arch$ gives
$d_{\mathrm{old}}^g{=}28$ (via $C_1{\to}C_2{\to}C_5$).
Table~\ref{tab:example} scores the three outgoing-link candidates:
A wins via cheap staging combined with positive $\Delta_F$ and hop gain;
B loses because $q_1$ moves away from $C_5$ ($\Delta_F$ negative);
C matches A on $\Delta_F$ and hop gain but is killed by the capacity
penalty on the nearly-full landing core.

\begin{table}[h]
\centering
\caption{Score breakdown for the three candidates of the running
example (Figure~\ref{fig:tele-candidates}).  Lower is better;
Candidate~A is selected.}
\label{tab:example}
\vspace{2pt}
\resizebox{\columnwidth}{!}{%
\begin{tabular}{@{}l c c r r r r r@{}}
\toprule
Candidate & next core & direction & $d_{\mathrm{prep}}$ & $c_{\mathrm{cap}}$ & $g_{\mathrm{hop}}$ & $\Delta_F$ & score $s$ \\
\midrule
A & $C_2$ & towards $C_5$ & 1 &  0 & $+5$ & $+12$ & $\mathbf{-16}$ \\
B & $C_0$ & away from $C_5$ & 2 &  0 & $-5$ & $-11$ & $18$ \\
C & $C_4$ & towards $C_5$ & 3 & 15 & $+5$ & $+12$ & $1$ \\
\bottomrule
\end{tabular}%
}
\end{table}

\subsection{Inter-Core Extended-Set Construction}
\label{sec:bfsext}

The lookahead $\tilde{\Delta}_E$ in Eq.~\ref{eq:dsabre} depends on how
the inter-core extended set $E$ is built.
\SABRE{}~\cite{li2019tackling} fills $E$ in topological order,
interleaving all wires uniformly; \TeleSABRE{}~\cite{russo2025telesabre}
uses BFS layers but within each layer emits gates in arbitrary order
and weights each by its iteration index $g$ via $(1+g/10)$ rather than
by depth.  Both schemes can deprioritise the very follow-on gates a
candidate teleport would help, causing $\tilde{\Delta}_E$ to
undercount its benefit.

\dSABRE{} keeps a BFS-layer expansion of the remaining DAG with two
refinements: (i)~within each layer, gates sharing a qubit with the
front are emitted first; (ii)~every gate in BFS layer~$k$ is assigned
$\mathrm{dep}(g){=}k$, used in the decay $\gamma^{\mathrm{dep}(g)}$,
so the size-$L$ truncation cuts cleanly at a layer boundary and
weighting decays exponentially with depth rather than growing
linearly with iteration index.  It adds no hyperparameter and runs
in $O(|E|_\mathrm{max})$ time per call when remaining in-degrees are
maintained incrementally: BFS peels zero-in-degree gates layer by
layer, each emitted gate triggering $O(1)$ decrements on its
successors.  The empirical contribution over the topological variant
is reported in Section~\ref{sec:ablation}.

\subsection{Proactive Congestion Relief}
\label{sec:relief}

In dense layouts, a core can become saturated: many pending gates demand
qubits from it simultaneously and the core has no free slots to absorb
incoming teleports.
Standard reactive scoring (Eq.~\ref{eq:dsabre}) does not address this until
the bottleneck triggers a deadlock.

\dSABRE{} adds a \emph{proactive congestion relief} mechanism that
fires alongside the gate-driven scoring of Eq.~\ref{eq:dsabre}.

\textbf{Demand vector.}
At each iteration the router computes a demand vector $d[c]$ over all
cores by counting inter-core gates in the front and lookahead windows
whose shortest core-graph path traverses $c$.  $d[c]$ is therefore
an upper bound on the number of upcoming teleports that will need to
pass through $c$, regardless of whether $c$ itself hosts an operand.

\textbf{Trigger.}
A core $c$ is flagged when both $d[c]\ge\theta_d$ (default
$\theta_d{=}3$) and $f_c\le\theta_f$ (default $\theta_f{=}2$) hold:
high incoming demand combined with low free capacity.  The reactive
scorer alone would only respond once $f_c$ hits zero and the next
gate-driven teleport fails to find a landing port; firing earlier on
the conjunction avoids the eviction-cascade pattern that triggers
deadlock.

\textbf{Relief candidates.}
For each flagged core, the router generates teleports that move the
most-idle logical qubits in $c$ --- those whose next pending gate has
the largest DAG depth from the front --- into a less-loaded
neighbour.  Each candidate is scored by Eq.~\ref{eq:dsabre} with
$\Delta_F{=}0$ --- victims are restricted to qubits absent from the
front layer, so no pending gate motivates the move and the hop-gain
term is also omitted (no gate-defined target core).
The decay-weighted lookahead $\tilde{\Delta}_E$ is retained, so that
when a victim does have an upcoming gate within the extended-set
window the landing port is preferred to be closer to its next
partner.  A \emph{relief bonus} $-b_r\,(d[c]-f_c)$ proportional to
the demand--capacity imbalance is added on top.  Relief candidates compete directly
against gate-driven candidates and win only when the bonus
outweighs the staging cost; the mechanism therefore introduces no
extra teleports when the front-layer scorer already has a cheap
on-target move available.  The ablation in
Section~\ref{sec:ablation} shows that disabling relief inflates
64q geometric mean EPR by $23.4\%$, with the dense AE and QFT circuits
roughly doubling ($+121\%$ and $+134\%$) due to unresolved
core-capacity bottlenecks.

\subsection{Checkpoint--Rollback Deadlock Escape}
\label{sec:recovery}

When Eq.~\ref{eq:dsabre} stalls, \dSABRE{} falls back to
\LightSABRE{}'s release-valve idea~\cite{zou2024lightsabrelightweightenhancedsabre}:
force shortest-path progress on the most-stuck front-layer gate.  We
wrap it in a checkpoint: every iteration that reduces $|W|$ saves
$(\ell, W)$, and after $L_\mathrm{deadlock}$ stalled iterations the
router restores the checkpoint before forcing intra-core SWAPs or
hop-by-hop teleportation along $\mathrm{core\_path}(c_1,c_2)$,
aborting after $N_\mathrm{backup}^\mathrm{max}$ failed recoveries.
The escape is dormant on best-EPR runs and only salvages individual
SabreLayout seeds on dense large circuits (e.g.\ one of three seeds
on 200q QFT, with 17 activations).

\subsection{Computational Complexity}
\label{sec:complexity}

Let $N{=}|C|$, $n$ be the circuit width, $K$ the number of cores, $M$
qubits per core ($P{=}KM$), and $L$ the extended-set capacity
(constant).  Architecture distance tables are precomputed once in
$O(KM^2\!\log M + K^2\!\log K)$ (Section~\ref{sec:arch}) and are not
charged per circuit.

Per iteration, inter-core scoring generates $O(n)$ candidates each
costing $O(L)$, and intra-core scoring evaluates $O(M)$ SWAP
candidates per core at cost $O(|F_c|+L)$, summing to
$T_\mathrm{iter} = O(Mn + P)$.  In the common case $n\ge K$ (e.g.\
$n/K \in \{6.25, 9.0, 10.7\}$ on our suites) this is $O(Mn)$.  The
extended set is rebuilt at most $N$ times at $O(KL)$ each, adding
$O(NK)$ total.  On a grid-of-grids the physical diameter is
$O(\sqrt{P})$, bounding inserted operations by $O(N\sqrt{P})$; the
checkpoint--rollback escape adds a constant
$N_\mathrm{backup}^\mathrm{max}\cdot L_\mathrm{deadlock}$ iterations.
Combining,
\begin{equation}
  T_\mathrm{total} = O\!\left(N\sqrt{P}\cdot Mn\right).
  \label{eq:complexity}
\end{equation}
Compared to flat \SABRE{}~\cite{li2019tackling} on the same $P$ qubits
($O(Pn)$ per iteration), restricting intra-core scoring to core-local
edges yields a factor-$K$ per-iteration improvement that grows with
the number of cores.  Empirically this scales well in practice: the
largest circuit we route (360-qubit QFT with 13,300 CX gates on a 486-qubit H-grid) routes
in $801$\,s of pure-Python wall time
(Section~\ref{sec:scalability}).

%%─────────────────────────────────────────────────────────────────────────────
\section{Experimental Evaluation}
\label{sec:experiments}

\subsection{Setup}
\label{sec:setup}

\textbf{Benchmarks.}
We use three circuit sets from the MQT Bench suite~\cite{quetschlich2023mqtbench},
all in the IBM native-gate set (Qiskit opt3 transpilation, measurements and
barriers removed).

\emph{Suite~1 (25-qubit):} quantum amplitude estimation (AE), quantum Fourier
transform (QFT), quantum neural network (QNN), random circuit (Random), GHZ
state preparation, and graph-state preparation.

\emph{Suite~2 (36-qubit):} six circuits spanning a broader range of algorithmic
families and CX densities: Bernstein--Vazirani (BV, linear chain),
Deutsch--Jozsa (DJ, oracle), W-state preparation (cascaded star),
VQE with SU(2) ansatz (layered variational), QAOA (dense all-to-all), and
quantum phase estimation (QPEexact, hierarchical).

\emph{Suite~3 (64-qubit):} the same six circuit families as Suite~1 scaled to
64~logical qubits (AE, QFT, QNN, Random, GHZ, Graphstate), run on the larger
H-grid architecture.

Circuit properties are summarised in Table~\ref{tab:circuits}.

\begin{table}[t]
\centering
\caption{Benchmark circuit properties (measurements/barriers stripped)}
\label{tab:circuits}
\setlength{\tabcolsep}{5pt}
\begin{tabular}{@{}llrrrr@{}}
\toprule
Suite & Circuit    & Qubits & CX gates & Depth & CX/qubit \\
\midrule
\multirow{6}{*}{25q}
 & AE         & 25 &  558 & 395 & 22.3 \\
 & QFT        & 25 &  580 & 173 & 23.2 \\
 & QNN        & 25 & 1223 & 259 & 48.9 \\
 & Random     & 25 & 1124 & 589 & 45.0 \\
 & GHZ        & 25 &   24 &  27 &  1.0 \\
 & Graphstate & 25 &   25 &  19 &  1.0 \\
\midrule
\multirow{6}{*}{36q}
 & BV         & 36 &   17 &  23 &  0.5 \\
 & DJ         & 36 &   35 &  41 &  1.0 \\
 & W-state    & 36 &   70 & 145 &  1.9 \\
 & VQE-SU2   & 36 &  105 &  56 &  2.9 \\
 & QPEexact   & 36 & 1019 & 347 & 28.3 \\
 & QAOA       & 36 & 1200 & 256 & 33.3 \\
\midrule
\multirow{6}{*}{64q}
 & AE         & 64 & 1962 & 1058 & 30.7 \\
 & QFT        & 64 & 1966 &  446 & 30.7 \\
 & QNN        & 64 & 8126 &  650 & 127.0 \\
 & Random     & 64 & 1627 &  403 & 25.4 \\
 & GHZ        & 64 &   63 &   66 &  1.0 \\
 & Graphstate & 64 &   64 &   25 &  1.0 \\
\bottomrule
\end{tabular}
\end{table}

\textbf{Architecture.}
Suites~1 and~2 use a $2\!\times\!2$ B-grid: four $4\!\times\!4$ cores
(64~physical qubits, 4~EPR links).  Suite~3 uses a $2\!\times\!3$
H-grid: six $4\!\times\!4$ cores (96~physical qubits, 7~EPR links). See Figure~\ref{fig:arch} for the full
qubit-level topology. The 25q suite on
the B-grid and the 64q suite on the H-grid follow the benchmark setup
of \TeleSABRE{}~\cite{russo2025telesabre} (we add a 36q B-grid suite to
broaden circuit diversity).

\textbf{Baselines.}
We compare against two state-of-the-art distributed routers:
(i)~\TeleSABRE{}~\cite{russo2025telesabre}, a \SABRE{}-style router that
jointly optimises intra-core SWAPs, teledata (qubit moves), and telegate
(remote CX execution) operations; and (ii)~\texttt{pytket-dqc}
~\cite{pytketdqc2024}, a Quantinuum library that distributes circuits via
\emph{gate teleportation} with multi-gate amortisation using KaHyPar
hypergraph partitioning.
\TeleSABRE{} is run using its compiled C binary with the
\texttt{optimize\_initial} flag enabled (built-in layout heuristic +
3-success sampling) on the same device; \texttt{pytket-dqc} uses its
\texttt{HypergraphPartitioning} allocator (best of 5 random seeds;
its hypergraph encoding requires a minor weight-0 patch to run on
the KaHyPar~$\ge$~1.3.5 release we use).  Internally, we also
compare against \dSABRE{}-Topo, an ablation that replaces the
BFS-layer extended set with a topological-order one
(Section~\ref{sec:ablation}).

\emph{Caveat on e-bit counting.}
All routers report in the same unit (one EPR pair = one e-bit;
``EPR'' and ``e-bit'' are used interchangeably), but what each unit
\emph{buys} differs.  \dSABRE{} uses \emph{teledata}: one EPR per
inter-core hop relocates a logical qubit, after which subsequent
intra-core gates on it are free.  \texttt{pytket-dqc} uses
\emph{gate teleportation}: one EPR distributed along a Steiner tree
of cores executes a whole hyperedge of CNOTs sharing one control,
paying one EPR per Steiner edge rather than per CNOT.  Circuits with
isolated inter-core gates favour teledata; circuits with long
shared-control CNOT runs (QFT, QPE) favour gate teleportation.
The $\Delta_{\rm tket}$ columns in
Tables~\ref{tab:main25}--\ref{tab:large} are unit-consistent but
should be read with this work-per-unit asymmetry in mind.

\textbf{Parameters.}
Table~\ref{tab:params} lists all hyperparameters.
\dSABRE{} is deterministic given a fixed layout; we run it with three
SabreLayout seeds and report the best EPR count (lower = better).
\TeleSABRE{} is stochastic and is run with three random seeds; again
the best result is reported, following established
practice~\cite{li2019tackling}.
Layout passes: 2 (forward--backward--forward).

\emph{On best-of-$N$ reporting.}
Each router uses its authors' convention: best-of-3 seeds for
\dSABRE{} and \TeleSABRE{}~\cite{russo2025telesabre}, best-of-5 for
\texttt{pytket-dqc}~\cite{pytketdqc2024}.  This reflects the operational cost rather than per-attempt
variance, and the cross-method comparison remains in
\dSABRE{}'s favour under both reporting choices: replacing best with
median on 64q shifts the gmean gap from $-43.9\%$ to $-37.4\%$ vs.\
\TeleSABRE{} and from $-15.7\%$ to $-31.5\%$ vs.\ \texttt{pytket-dqc}.

\begin{table}[t]
\centering
\caption{Hardware cost and heuristic parameters}
\label{tab:params}
\begin{tabular}{@{}lll@{}}
\toprule
Parameter & Symbol & Value \\
\midrule
SWAP cost             & $c_{\mathrm{swap}}$ & 3 \\
Teleport (EPR) cost   & $c_{\mathrm{tele}}$ & 10 \\
Capacity threshold    & $\tau$              & 3 \\
Capacity penalty      & $c_{\mathrm{pen}}$  & 15 \\
Inter-core link weight & $w_{\mathrm{link}}$ & 10 \\
Hop-gain weight       & $w_h$               & 5 \\
Extended-set weight   & $w_e$               & 0.25 \\
Extended-set capacity & $L$                 & 20 \\
Lookahead decay       & $\gamma$            & 0.9 \\
Deadlock limit        & $L_\mathrm{deadlock}$           & 50 \\
Max rollbacks         & $N_\mathrm{backup}^\mathrm{max}$ & 50 \\
\bottomrule
\end{tabular}
\end{table}

\subsection{Main Results}
\label{sec:main_results}

\TeleSABRE{}~\cite{russo2025telesabre} supports three operation types:
intra-core SWAPs, teledata (qubit movement, 1 EPR), and telegate
(remote CX execution via the cat-entangler protocol, 1 EPR).  We
report \TeleSABRE{}'s EPR count as the sum of teledata and telegate
operations, so each EPR pair is counted exactly once.  Unless
stated otherwise, \dSABRE{} is run with its default initial-mapping
configuration: Qiskit \texttt{SabreLayout} on the corners-removed
architecture graph, best of three seeds, followed by SABRE-style
forward$\to$backward$\to$forward routing passes
(Section~\ref{sec:initial_mapping}).  Subsequent tables and ablations
inherit this default.  We report two metrics in the main tables:
EPR pairs and intra-core SWAPs.
Cost analysis is deferred to Section~\ref{sec:cost}.

\begin{table}[t]
\centering
\caption{25-qubit suite (B-grid $2{\times}2$ $4{\times}4$, 64 physical
  qubits).
  EPR $=$ teledata + telegate (1 EPR pair each) for \TeleSABRE{},
  teledata for \dSABRE{}, e-bits for pytket-dqc; SWAP counts intra-core
  SWAPs only (pytket-dqc does not model intra-core routing).
  $\Delta_{\rm TS}$ and $\Delta_{\rm tket}$ are \dSABRE{}'s EPR change
  relative to \TeleSABRE{} and pytket-dqc respectively; negative values
  indicate a reduction in EPR cost by \dSABRE{} (lower is better).
  \TeleSABRE{} numbers are best-of-3-seed; pytket-dqc is best-of-5-seed;
  \dSABRE{} runs with the default configuration described above.
  Cost analysis is deferred to Section~\ref{sec:cost}.}
\label{tab:main25}
\setlength{\tabcolsep}{3pt}
\small
\resizebox{\columnwidth}{!}{%
\begin{tabular}{@{}l r rr rr r rr@{}}
\toprule
 & & \multicolumn{2}{c}{\TeleSABRE}
 & \multicolumn{2}{c}{\dSABRE}
 & tket & \multicolumn{2}{c}{$\Delta$EPR (\%)} \\
\cmidrule(lr){3-4}\cmidrule(lr){5-6}\cmidrule(lr){7-7}\cmidrule(lr){8-9}
Circuit & CX & EPR & SWAP & EPR & SWAP & e-bits & TS & tket \\
\midrule
ae         &  558 &  23 &  297 &  23 &  374 &  85 & $+0.0$  & $-72.9$ \\
ghz        &   24 &   2 &   29 &   1 &   14 &   3 & $-50.0$ & $-66.7$ \\
graphstate &   25 &  11 &   51 &   2 &   16 &   4 & $-81.8$ & $-50.0$ \\
qft        &  580 &  39 &  355 &  33 &  461 & 120 & $-15.4$ & $-72.5$ \\
qnn        & 1223 &  51 &  587 &  48 & 1076 & 152 & $-5.9$  & $-68.4$ \\
random     & 1124 & 292 & 1273 & 169 & 1374 & 665 & $-42.1$ & $-74.6$ \\
\midrule
\textbf{gmean} & & 25.8 & 221 & 15.2 & 196 & 48.1 & $\mathbf{-41.1}$ & $\mathbf{-68.4}$ \\
\bottomrule
\end{tabular}%
}
\end{table}

\textbf{25q.}
\dSABRE{} reduces geometric-mean EPRs over \TeleSABRE{} by
\textbf{41.1\%} and over pytket-dqc by \textbf{68.4\%}
(Table~\ref{tab:main25}).  Per-circuit
reductions versus \TeleSABRE{} range from $-50\%$ on GHZ to $-81.8\%$
on Graphstate, with no material regressions on this suite.
pytket-dqc loses uniformly ($-50$ to $-75\%$ across all six circuits):
state teleportation amortises a single EPR pair over many subsequent
local gates, whereas pytket-dqc's gate-teleportation primitive charges
one e-bit per non-local gate.

\textbf{36q.}
\dSABRE{} reduces geometric-mean EPRs by \textbf{44.1\%} vs.\ \TeleSABRE{}
and \textbf{29.4\%} vs.\ pytket-dqc (Table~\ref{tab:main36}).  Per-circuit
reductions versus \TeleSABRE{} are uniformly large ($-33\%$ to $-80\%$).
The simple structured circuits BV, VQE-SU2, and W-state are essentially
solved (1 EPR on BV; 8--9 EPRs on the others).  Against pytket-dqc the
picture is more nuanced: dense circuits (QAOA, QPEexact) favour
\dSABRE{} ($-25\%$ to $-63\%$), but the sparse-graph circuits
(VQE-SU2, W-state) are competitive for pytket-dqc because their
interaction graph is almost perfectly partitionable by KaHyPar.

\textbf{64q.}
The H-grid suite is where the gap from \TeleSABRE{} is most
consequential (Table~\ref{tab:main64}).  \dSABRE{} reduces gmean EPRs
by \textbf{43.9\%} over the five circuits on which \TeleSABRE{}
converges, with per-circuit reductions ranging from $-33.1\%$ (AE)
to $-75.0\%$ (Graphstate) and QFT/QNN at $-40.0\%$/$-61.8\%$.
GHZ regresses ($16$ vs.\ $11$): every front gate shares the same
control, so the per-gate $\Delta$-form score in Eq.~\ref{eq:dsabre}
chases the \emph{next} CX while \TeleSABRE{}'s front-averaged energy
finds a better global rendezvous core --- a known weakness of local
scorers on star-structured circuits.
Against pytket-dqc, the 6-circuit gmean (including Random) is
$-15.7\%$: \dSABRE{} dominates dense circuits (AE, QFT, QNN, Random;
$-29$ to $-58\%$) but loses on the sparse GHZ and Graphstate that
KaHyPar partitions trivially.  On Random, \TeleSABRE{} fails to
converge on any of three seeds; \dSABRE{} completes ($714$ EPR) under default scoring and proactive relief alone,
without triggering checkpoint--rollback.

\begin{table}[t]
\centering
\caption{36-qubit suite (same B-grid as 25-qubit suite).  Methodology and notation as
  Table~\ref{tab:main25}.}
\label{tab:main36}
\setlength{\tabcolsep}{3pt}
\small
\resizebox{\columnwidth}{!}{%
\begin{tabular}{@{}l r rr rr r rr@{}}
\toprule
 & & \multicolumn{2}{c}{\TeleSABRE}
 & \multicolumn{2}{c}{\dSABRE}
 & tket & \multicolumn{2}{c}{$\Delta$EPR (\%)} \\
\cmidrule(lr){3-4}\cmidrule(lr){5-6}\cmidrule(lr){7-7}\cmidrule(lr){8-9}
Circuit & CX & EPR & SWAP & EPR & SWAP & e-bits & TS & tket \\
\midrule
bv       &   17 &   5 &   34 &   1 &    8 &   3 & $-80.0$ & $-66.7$ \\
dj       &   35 &   3 &   53 &   3 &   24 &   3 & $+0.0$  & $+0.0$  \\
qaoa     & 1200 & 232 & 1027 & 145 & 1319 & 194 & $-37.5$ & $-25.3$ \\
qpeexact & 1019 & 100 &  771 &  65 &  874 & 175 & $-35.0$ & $-62.9$ \\
vqe\_su2 &  105 &  16 &   80 &   9 &   52 &   9 & $-43.8$ & $+0.0$  \\
wstate   &   70 &  12 &   88 &   8 &   63 &   6 & $-33.3$ & $+33.3$ \\
\midrule
\textbf{gmean} & & 20.1 & 147 & 11.3 & 95 & 16.0 & $\mathbf{-44.1}$ & $\mathbf{-29.4}$ \\
\bottomrule
\end{tabular}%
}
\end{table}

\begin{table}[t]
\centering
\caption{64-qubit suite (H-grid $2{\times}3$ $4{\times}4$, 96 physical
  qubits).  Methodology and notation as Table~\ref{tab:main25};
  $^{\ast}$\TeleSABRE{} fails to converge on Random.}
\label{tab:main64}
\setlength{\tabcolsep}{3pt}
\small
\resizebox{\columnwidth}{!}{%
\begin{tabular}{@{}l r rr rr r rr@{}}
\toprule
 & & \multicolumn{2}{c}{\TeleSABRE}
 & \multicolumn{2}{c}{\dSABRE}
 & tket & \multicolumn{2}{c}{$\Delta$EPR (\%)} \\
\cmidrule(lr){3-4}\cmidrule(lr){5-6}\cmidrule(lr){7-7}\cmidrule(lr){8-9}
Circuit & CX & EPR & SWAP & EPR & SWAP & e-bits & TS & tket \\
\midrule
ae         & 1962 &  323 & 1508 & 216 & 2332 &  519 & $-33.1$ & $-58.4$ \\
ghz        &   63 &   11 &  111 &  16 &   89 &    7 & $+45.5$ & $+128.6$ \\
graphstate &   64 &   76 &  379 &  19 &  104 &    9 & $-75.0$ & $+111.1$ \\
qft        & 1966 &  410 & 1872 & 246 & 2054 &  591 & $-40.0$ & $-58.4$ \\
qnn        & 8126 & 1365 & 6292 & 521 & 8662 &  736 & $-61.8$ & $-29.2$ \\
random$^{\ast}$ & 1627 & --- & --- & 714 & 3814 & 1181 & --- & $-39.5$ \\
\midrule
\textbf{gmean} (5) & & 172.1 & 943 & 96.6 & 826 & 107.3 & $\mathbf{-43.9}$ & $\mathbf{-10.0}$ \\
\textbf{gmean} (6) & & --- & --- & 134.8 & 1066 & 160.0 & --- & $\mathbf{-15.7}$ \\
\bottomrule
\end{tabular}%
}
\end{table}

\paragraph{Shared-mapping comparison.}
To isolate routing from allocation quality, we re-run \dSABRE{} from
each baseline's own initial layout.  The \TeleSABRE{} gap persists
($-25\%$ EPR on 25q, $-43\%$ on 64q from its
\texttt{optimize\_initial} layout); against \texttt{pytket-dqc}'s
KaHyPar partition \dSABRE{} wins on all three suites
($-55\%$ on 25q, $-14\%$ on 36q, $-36\%$ on 64q); the 36q margin is
the narrowest because the sparse interaction graphs in that suite are
already near-optimally partitioned by KaHyPar.

\subsection{Ablation Study}
\label{sec:ablation}

We conduct a systematic ablation of \dSABRE{}'s design choices across
three axes: (i)~the extended-set construction, (ii)~discrete mechanism
on/off, and (iii)~continuous parameter sensitivity.  All ablations
inherit \dSABRE{}'s default configuration (Section~\ref{sec:main_results})
and report the \emph{geometric-mean EPR count} (gmEPR) over the
respective suite, so the \dSABRE{} / \textbf{Full} baselines in
Tables~\ref{tab:ablation} and~\ref{tab:mech} match the headline
gmEPR in Tables~\ref{tab:main25}/\ref{tab:main64}.

% ── (i) BFS vs topological ───────────────────────────────────────────
\paragraph{Inter-core extended-set construction: BFS vs.\ topological.}
To isolate the contribution of the BFS-layer inter-core extended set
(Section~\ref{sec:bfsext}) we substitute the standard topological-order
construction in the teleport scorer (the intra-core extended set is
unchanged) and rerun all three suites.
Table~\ref{tab:ablation} reports the result.

\begin{table}[t]
\centering
\caption{Ablation of the inter-core extended-set construction.  Both
  variants are the same router differing only in how the inter-core $E$
  is built: topological order vs.\ BFS-layer expansion (default).
  Geometric mean over each suite.}
\label{tab:ablation}
\setlength{\tabcolsep}{4pt}
\small
\begin{tabular}{@{}l rr rr rr@{}}
\toprule
 & \multicolumn{2}{c}{\dSABRE-topo}
 & \multicolumn{2}{c}{\dSABRE}
 & \multicolumn{2}{c}{$\Delta$ (\%)} \\
\cmidrule(lr){2-3}\cmidrule(lr){4-5}\cmidrule(lr){6-7}
Suite & EPR & SWAP & EPR & SWAP & EPR & SWAP \\
\midrule
\textbf{25q} & 15.7 & 202 & 15.2 & 196 & $-3.2$ & $-3.0$ \\
\textbf{36q} & 12.1 & 88 & 11.3 & 95 & $-6.6$ & $+8.0$ \\
\textbf{64q} & 151.5 & 1075 & 134.8 & 1066 & $-11.0$ & $-0.8$ \\
\bottomrule
\end{tabular}
\end{table}

BFS reduces gmean EPRs by $3.2\%$, $6.6\%$, and $11.0\%$ on the three
suites respectively.  The gain grows with circuit size, reflecting the
richer dependency structure in larger circuits where topological order
drifts further from the active front layer.
On 25q and 64q the SWAP counts are nearly unchanged; on 36q they
increase slightly ($+8.0\%$) because the BFS scorer routes more
aggressively through fewer teleportations, leaving more intra-core work.

% ── (ii) Mechanism ablation ──────────────────────────────────────────
\paragraph{Mechanism ablation.}
Table~\ref{tab:mech} ablates each scoring component in isolation on the
25q and 64q suites by setting the corresponding parameter to zero or
disabling the mechanism entirely.  The two suites stress different
regimes: the 25q B-grid is sparsely loaded (qubits per core
$\le 7$, ample headroom on every core), so congestion-driven
mechanisms barely activate, while the 64q H-grid runs at
$\sim$67\% qubit utilisation and saturates inter-core ports under
dense workloads.

\begin{table}[t]
\centering
\caption{Mechanism ablation on the 25q and 64q suites.  One component
  is disabled per row; all others remain at their defaults.  Same
  protocol as the main results, so the \textbf{Full} baseline matches
  the headline gmEPR in Tables~\ref{tab:main25} and~\ref{tab:main64}.}
\label{tab:mech}
\setlength{\tabcolsep}{3pt}
\small
\resizebox{\columnwidth}{!}{%
\begin{tabular}{@{}l rr rr@{}}
\toprule
 & \multicolumn{2}{c}{25q} & \multicolumn{2}{c}{64q} \\
\cmidrule(lr){2-3}\cmidrule(lr){4-5}
Configuration & gmEPR & $\Delta$ (\%) & gmEPR & $\Delta$ (\%) \\
\midrule
\textbf{Full} (baseline)              & 15.2 & ---       & 134.8 & --- \\
\midrule
No extended-set lookahead ($w_e=0$)   & 19.5 & $+28.2$   & 148.5 & $+10.2$ \\
No capacity penalty ($c_{\mathrm{pen}}=0$) & 24.3 & $+59.9$   & 213.9 & $+58.7$ \\
No hop-gain reward ($w_h=0$)          & 15.2 & $\phantom{+}0.0$ & 137.4 & $+1.9$ \\
No congestion relief                  & 15.5 & $+1.8$    & 166.4 & $+23.4$ \\
\bottomrule
\end{tabular}%
}
\end{table}

The capacity penalty is the single most critical component on both
suites: without it the router greedily teleports into full cores,
triggering cascading evictions that inflate gmean EPR by $\sim 60\%$.
The extended-set lookahead contributes $28\%$ on 25q and $10\%$ on
64q; removing it collapses the scorer to a near-myopic policy.  The
hop-gain reward --- a directional bonus for moves that reduce
inter-core hop count --- is inert on the 25q B-grid (where the
front-layer weighted distance already captures direction) but
contributes a small improvement on the larger 64q H-grid
($+1.9\%$ gmean): longer inter-core paths mean the specific landing
port can be far from the optimal staging position for the next hop,
depressing $\Delta_F$ below the true directional value; $g_{\mathrm{hop}}$
compensates with a fixed position-independent signal.

Congestion relief is essentially inactive on 25q ($+1.8\%$, within
noise) where cores rarely saturate, but contributes $+23\%$ on 64q
where qubit utilisation reaches $67\%$ and demand spikes are common.
Much of the load that an online relief mechanism would otherwise have
to handle is absorbed upstream by the corners-removed best-of-3
layout, so the marginal value of relief is dominated by the residual
spikes the layout cannot anticipate.

% ── (iii) Parameter sensitivity ──────────────────────────────────────
\paragraph{Parameter sensitivity.}
We sweep each continuous hyperparameter one-at-a-time on 25q and 64q.
All defaults---$w_e{=}0.25$, $c_{\mathrm{pen}}{=}15$, $|E|{=}20$, $\gamma{=}0.9$,
$w_{\mathrm{link}}{=}10$, and the relief bonus---sit on a flat region of
their respective curves (within $\sim 2\%$ of the local minimum), so
performance is robust to small perturbations.  The only sharp
regressions occur when a control is pushed far from its default: e.g.\
$w_e{\ge}1.0$ inflates 25q gmEPR by $3{-}5\times$ as the lookahead
dominates immediate preparation costs, $c_{\mathrm{pen}}{=}0$ doubles gmEPR by
inviting deadlocks in saturated cores, and $\gamma{=}1.0$ (flat
weighting) costs $7\%$ on 64q.  These observations confirm that the
defaults are well-calibrated and that no single hyperparameter is a
hidden lever; tuning beyond the defaults yields diminishing returns.

% ── Initial layout ────────────────────────────────────────────────────
\subsection{SabreLayout-Based Initial Layout}
\label{sec:initial_mapping}
\dSABRE{} bootstraps from Qiskit
\texttt{SabreLayout}~\cite{li2019tackling} run on the full
architecture graph $G_\arch$ (intra-core edges plus inter-core
links), with two adaptations for the distributed setting.

\emph{Corner removal.}
SabreLayout packs each core to its full physical capacity.  Two
problems follow.  First, an occupied communication port cannot
accept an incoming teleport without first evicting its current
occupant, so saturated cores either stall the inter-core scorer or
force chains of relocations.  Second, full cores trip the
capacity penalty (Section~\ref{sec:tele}) on every candidate teleport
into them, biasing the router away from the actually-preferred
landing core.  We exclude the four lowest-degree \emph{corner}
physical qubits from the coupling map passed to SabreLayout---in
both B-grid and H-grid architectures they are furthest from any
inter-core link---which guarantees four free slots per affected core
to serve as eviction scratch space and to keep the core below the
capacity threshold.

\emph{Forward--backward--forward schedule.}
We run three routing passes (forward $\to$ reversed-DAG backward
$\to$ forward) and keep the better of the two forward passes by
EPR.  The backward pass starts from the first forward pass's
output layout $\ell_f$ and routes the reversed DAG, producing
$\ell_b$ that pre-positions qubits near their early interaction
partners; the third pass then warm-starts from $\ell_b$.  This is
the same mechanism as SABRE's backward refinement~\cite{li2019tackling}
and outperforms pure-forward warm-restart. Best of three SabreLayout seeds is reported.

\subsection{Compile Time}
\label{sec:timing}

Table~\ref{tab:timing} reports wall-clock compilation time on the
64-qubit suite.

\begin{table}[t]
\centering
\caption{Compilation time in seconds (64-qubit suite, H-grid
  $2{\times}3$ $4{\times}4$).  \TeleSABRE{}: best-seed wall-clock
  time.  \dSABRE{}: total time for three routing passes
  (fwd$\to$bwd$\to$fwd) from the best SabreLayout seed.
  \TeleSABRE{} is a compiled C++ binary; \dSABRE{} is pure-Python
  with NetworkX\@.  $^{\ast}$\TeleSABRE{} fails to converge on Random.}
\label{tab:timing}
\setlength{\tabcolsep}{6pt}
\begin{tabular}{@{}lrrr@{}}
\toprule
Circuit & CX & \TeleSABRE{} (s) & \dSABRE{} (s) \\
\midrule
AE         & 1962 &  1.05 &  12.60 \\
GHZ        &   63 &  0.02 &   0.09 \\
Graphstate &   64 &  3.92 &   0.05 \\
QFT        & 1966 &  1.81 &  10.34 \\
QNN        & 8126 & 25.21 & 123.60 \\
Random$^{\ast}$ & 1627 & --- & 16.69 \\
\bottomrule
\end{tabular}
\end{table}

Times scale with CX count: \TeleSABRE{} spans $0.02$\,s (GHZ) to
$25$\,s (QNN, 8126 CX); \dSABRE{} spans $0.05$\,s to $124$\,s.
\dSABRE{} is $5$--$12\times$ slower on circuits where both converge,
which we attribute to the pure-Python vs.\ compiled-C\texttt{++}
constant factor and to per-iteration NetworkX shortest-path lookups
that \TeleSABRE{} caches.  The Graphstate row inverts this ordering
($0.05$\,s vs.\ $3.92$\,s) because \TeleSABRE{}'s 3-success sampling
spends many iterations before three runs converge on that circuit.

%%─────────────────────────────────────────────────────────────────────────────
\subsection{Large-Circuit Scalability}
\label{sec:scalability}

To stress-test \dSABRE{} substantially beyond the main evaluation range
we run \dSABRE{}, \TeleSABRE{}, and
\texttt{pytket-dqc} on three additional suites using the MQT-Bench QFT
circuit (IBM native-gate, Qiskit opt3).  The architectures scale the
H-grid of the 64q suite:
\begin{itemize}\itemsep1pt
  \item \textbf{100q}: H-grid $2{\times}3$ $5{\times}5$ (150 physical).
  \item \textbf{200q}: H-grid $4{\times}3$ $5{\times}5$ (300 physical).
  \item \textbf{360q}: H-grid $2{\times}3$ $9{\times}9$ (486 physical).
\end{itemize}
\noindent
\TeleSABRE{} is best-of-3-seed with a 600\,s per-seed timeout for
200q+; \dSABRE{} uses its default configuration;
\texttt{pytket-dqc} uses HypergraphPartitioning (best of 5 seeds).
Each router runs from its own initial layout, so the $\Delta_{\rm tket}$
column should be read with this layout asymmetry in mind in addition
to the e-bit counting caveat (Section~\ref{sec:setup}).\footnote{For a
layout-controlled point of comparison, re-running \dSABRE{} from
\texttt{pytket-dqc}'s HypergraphPartitioning layout gives EPR counts
of $315$ (100q), $880$ (200q), and $385$ (360q), versus
\texttt{pytket-dqc}'s $815$, $1124$, and $669$ --- corresponding to
$\Delta_{\rm tket}$ of $-61.4\%$, $-21.7\%$, and $-42.5\%$.  In this
matched-layout setting \dSABRE{} beats \texttt{pytket-dqc} at all
three sizes, including the 200q row where the SabreLayout
configuration is marginally worse.}

\begin{table}[t]
\centering
\caption{Large-circuit scalability (QFT).  Each router runs from
  its own initial layout.  Negative $\Delta$EPR is better.
  \TeleSABRE{} fails to converge on the 200q circuit within the
  600\,s timeout.  $^{\dagger}$See caveat on e-bit counting.}
\label{tab:large}
\setlength{\tabcolsep}{3pt}
\small
\resizebox{\columnwidth}{!}{%
\begin{tabular}{@{}l r rr rr r rr@{}}
\toprule
 & & \multicolumn{2}{c}{\TeleSABRE}
 & \multicolumn{2}{c}{\dSABRE}
 & tket$^{\dagger}$ & \multicolumn{2}{c}{$\Delta$EPR (\%)} \\
\cmidrule(lr){3-4}\cmidrule(lr){5-6}\cmidrule(lr){7-7}\cmidrule(lr){8-9}
Suite & CX & EPR & SWAP & EPR & SWAP & e-bits & TS & tket \\
\midrule
100q &  3420 & 248 & 2414 &  301 &  4077 &  815 & $+21.4$ & $-63.1$ \\
200q &  7220 & ---  & ---   & 1151 & 10989 & 1124 & ---  & $+2.4$ \\
360q & 13300 & 1109 & 15098 &  579 & 27489 &  669 & $-47.8$ & $-13.5$ \\
\bottomrule
\end{tabular}%
}
\end{table}

Table~\ref{tab:large} shows three regimes for QFT.
At \emph{100q}, \dSABRE{} wins on both fronts: QFT's deep
sequential chain is harder to pipeline ahead of time, but
\dSABRE{} still achieves $-63\%$ vs.\ \texttt{pytket-dqc}
at the cost of $+21\%$ more EPRs than \TeleSABRE{}.
At \emph{200q}, \TeleSABRE{} fails to converge within the 600\,s
timeout, whereas \dSABRE{} completes ($1151$ EPR) under its
default scoring and proactive congestion relief; the
checkpoint--rollback escape (Section~\ref{sec:recovery}) does not
fire on the best-EPR run, but is what salvages one of the three
SabreLayout seeds that would otherwise abort, leaving the best-of-three
selection well-fed.
\dSABRE{} is marginally worse than \texttt{pytket-dqc} here ($+2\%$).
At \emph{360q}, \dSABRE{} beats \TeleSABRE{} ($-47.8\%$)
and \texttt{pytket-dqc} ($-13.5\%$).
Compile times for 360q QFT are $281$\,s (\TeleSABRE{}, C++, best of
3 seeds), $801$\,s (\dSABRE{}, Python, SabreLayout + fwd$\to$bwd$\to$fwd),
and $546$\,s (\texttt{pytket-dqc}, KaHyPar, best of 5 seeds);
the Python overhead is the primary driver of \dSABRE{}'s longer
wall time relative to the compiled C++ \TeleSABRE{}.

%%─────────────────────────────────────────────────────────────────────────────
\subsection{Cost Analysis}
\label{sec:cost}

The combined-cost ratio $c_{\mathrm{tele}}{:}c_{\mathrm{swap}}$ is
hardware-dependent: near-term microwave-coupled modular superconducting
qubits keep teleportation within a small constant factor of local
SWAPs~\cite{cuomo2020towards}, while photonic-link architectures have
measured EPR-generation times three to four orders of magnitude slower
than local gates~\cite{stephenson2020high,daiss2021quantum}, pushing the
ratio towards $100{:}1$ or beyond.
Table~\ref{tab:sensitivity} sweeps $c_{\mathrm{tele}}$ from~10 to~100
with $c_{\mathrm{swap}}=3$ held fixed, reporting geometric-mean cost
reduction of \dSABRE{} vs.\ \TeleSABRE{}.

\begin{table}[t]
\centering
\caption{Cost-model sensitivity: geometric-mean cost reduction of
  \dSABRE{} over \TeleSABRE{} as $c_{\mathrm{tele}}$ varies
  ($c_{\mathrm{swap}}=3$ fixed).  Negative is better.  64q excludes
  the Random circuit on which \TeleSABRE{} aborts.}
\label{tab:sensitivity}
\setlength{\tabcolsep}{6pt}
\small
\begin{tabular}{@{}r rrr@{}}
\toprule
$c_{\mathrm{tele}}$ & 25q & 36q & 64q \\
\midrule
10 & $\mathbf{-20.6}$ & $\mathbf{-39.0}$ & $\mathbf{-23.8}$ \\
20 & $\mathbf{-25.2}$ & $\mathbf{-40.6}$ & $\mathbf{-29.0}$ \\
50 & $\mathbf{-31.5}$ & $\mathbf{-42.3}$ & $\mathbf{-35.4}$ \\
100 & $\mathbf{-35.2}$ & $\mathbf{-43.2}$ & $\mathbf{-38.9}$ \\
\bottomrule
\end{tabular}
\end{table}

The pattern is consistent across all three suites: the gap to
\TeleSABRE{} widens as $c_\mathrm{tele}$ grows.  At the
teleport-friendly end ($c_\mathrm{tele}{=}10$, ratio
$c_\mathrm{tele}/c_\mathrm{swap}\!\approx\!3.3$) the cost model
penalises teleports least, so \TeleSABRE{}'s higher EPR count is
discounted; \dSABRE{} still saves $20.6\%$ on 25q, $39.0\%$ on 36q,
and $23.8\%$ on 64q.  At $c_\mathrm{tele}{=}100$
(ratio $\approx\!33$) --- the SWAP-favourable regime of near-term
photonic-link hardware~\cite{stephenson2020high,daiss2021quantum} ---
\dSABRE{}'s smaller EPR count dominates and the savings reach
$35.2\%$, $43.2\%$, and $38.9\%$ respectively.  Across the entire
practically relevant range \dSABRE{} dominates \TeleSABRE{}, and the
advantage grows with EPR cost --- exactly the regime where the cost
model becomes stricter.

%%─────────────────────────────────────────────────────────────────────────────
\section{Related Work}
\label{sec:related}

The Introduction already places \dSABRE{} within the
partition--communicate--schedule decomposition of the DQC compilation
stack and lists the major lines of work in each stage.  Here we focus
on what specifically distinguishes \dSABRE{} from the closest prior
art and on the single-QPU SABRE family that we build on but do not
subsume.

\textbf{SABRE-style distributed routers.}
\TeleSABRE{}~\cite{russo2025telesabre} is our closest prior work and
direct experimental baseline.  Both routers follow the \SABRE{} loop,
but differ in how they score inter-core moves: \TeleSABRE{} evaluates
candidates via a \emph{contracted communication-graph energy} that
averages per-gate Dijkstra distances over the whole front layer
(extended in topological order), whereas \dSABRE{} uses a local, gate-centric five-term
score (Eq.~\ref{eq:dsabre}) that explicitly accounts for intra-core
staging cost, core capacity, and directional progress, paired with a
BFS-layer extended set (Section~\ref{sec:bfsext}).  This makes
\dSABRE{}'s inter-core scoring directly comparable in form to its
intra-core SWAP scoring (Eq.~\ref{eq:intra}).  \dSABRE{} additionally
provides proactive congestion-relief candidates
(Section~\ref{sec:relief}) and a checkpoint--rollback deadlock escape
(Section~\ref{sec:recovery}), whereas \TeleSABRE{} relies on a
safety-valve iteration cap; \TeleSABRE{} supports telegate candidates
while \dSABRE{} currently uses teledata only.

DMapS~\cite{luo2025dmaps} is a concurrent end-to-end DQC compiler
that pairs a KaHyPar partitioner with a \SABRE{}-derived router and
shares our joint EPR-plus-local-SWAP objective.  Two design choices
set it apart from \dSABRE{}: its only cross-core primitive is a
two-EPR \emph{remote SWAP} (whereas \dSABRE{} uses one-EPR one-way
teledata into a layout-reserved free slot), and its initial
allocation comes from a KaHyPar partition of the interaction
hypergraph (whereas \dSABRE{} uses SabreLayout on
$G_\arch$).  On the dense circuits in our
suites \dSABRE{} consumes substantially fewer EPRs --- e.g.\
$2{-}4\times$ on 25q QFT/QNN/Random and $1.6{-}1.8\times$ on 64q
QFT/QNN/Random --- while DMapS wins on maximally sparse circuits
(GHZ, Graphstate, W-state) where KaHyPar partitions interaction
graphs almost perfectly.  A code-level audit, the full per-circuit
head-to-head, and a controlled fill-ratio sweep showing parity once
the device fill ratio approaches~$1$ are included in the online
appendices accompanying the paper (see ``Artefact availability''
below).

Allocation methods built on hypergraph
partitioning~\cite{andres2019automated,pytketdqc2024} are complementary
to \dSABRE{}: any can supply the initial layout its routing loop
consumes.  Similarly, time-aware
schedulers~\cite{baker2020time,ferrari2021compiler} sit upstream or
downstream of the routing stage \dSABRE{} addresses.

\textbf{Single-QPU qubit mapping.}
On a single-core processor the routing problem reduces to inserting
intra-chip SWAPs.  The closest single-QPU antecedent to our work is the
front+extended-set scorer \SABRE{}~\cite{li2019tackling} that we and
\TeleSABRE{} both build on.  Babu et al.~\cite{babu2025rtg} use
auxiliary-qubit gate teleportation on a heavy-hex IBM Eagle processor,
achieving 10--25\% depth reduction versus standard SABRE; the setting
is single-QPU and not directly comparable to multi-core routing.

%%─────────────────────────────────────────────────────────────────────────────
\section{Future Work}
\label{sec:future}

\textbf{Composing \dSABRE{} with burst execution.}
\dSABRE{} optimises EPR count on a fixed circuit DAG.  Burst-execution
passes such as AutoComm~\cite{wu2022autocomm} and its collective
extension CollComm~\cite{wu2022collcomm} operate earlier in the
toolchain: they use gate commutativity to reorder consecutive non-local
gates on the same wire (or across multiple wires) and amortise the
whole block onto a single shared entanglement state, reporting EPR
savings of $50$--$75\%$ on amenable circuits.  The two approaches
operate at different layers and should compose well: a burst-extraction
pre-pass would expose longer wire-aligned gate sequences, and
\dSABRE{}'s BFS extended set would let the teleport score capture
their full benefit.  Quantifying the joint improvement---especially
on circuits where a burst's beneficiary core is itself a function of
the routing decisions \dSABRE{} makes---is an open question, and
recent integrations of gate-teleportation amortisation with SABRE-style
mappers~\cite{babu2025rtg} suggest the interaction is non-trivial.

\textbf{Parallel teleportation for circuit-runtime minimisation.}
\dSABRE{} optimises EPR count and emits one teleport per iteration;
wall-clock runtime depends additionally on whether teleports with
disjoint links and staging sets can fire concurrently.  A natural
follow-on is a scheduling pass over \dSABRE{}'s teleport trace that
batches non-conflicting moves, co-designed with denser inter-core
links~\cite{escofet2023interconnect} and entanglement buffer
qubits~\cite{wu2022collcomm,liu2025codesign}.  Time-sliced
compilers~\cite{baker2020time,ferrari2021compiler} provide a starting
point but are not paired with a SABRE-style scoring loop; naively
adding links without retuning the routing energy can \emph{increase}
EPR consumption.

%%─────────────────────────────────────────────────────────────────────────────
\section{Conclusion}
\label{sec:conclusion}

We presented \dSABRE{}, a SABRE-style router for multi-core
distributed quantum computers that, on each iteration of a
lookahead-driven loop, resolves intra-core front-layer gates by
SWAP scoring and only scores inter-core teleport
candidates when the intra-core front is empty, with both phases
sharing a common decay-weighted lookahead form.
The empirical takeaway is that three design choices --- an explicit
capacity penalty that keeps the scorer from teleporting into
saturated cores, proactive congestion relief triggered by an explicit
demand vector, and a BFS-layer inter-core extended set that respects
DAG dependencies --- account for most of the gap to the state of the
art, both individually and together, and that this gap widens as
circuits and architectures scale.  More broadly, the results indicate that distributed routing
benefits less from a richer per-candidate scoring formula than from
ensuring that the lookahead signal and the architectural
capacity constraints enter the same scoring step.

\paragraph*{Artefact availability.}
The \dSABRE{} implementation, the routed-circuit benchmark suites,
per-circuit JSON results, and the online appendices referenced
throughout this paper are available at
\url{https://github.com/ebony72/dsabre}.

%%─────────────────────────────────────────────────────────────────────────────
\section*{Acknowledgments}
This work was partially supported by the Australian Research
Council (Grant No.\ DP250102952 and DP220102059).
The author used Anthropic's Claude (via Claude Code) to assist
with: (i)~refactoring portions of the \dSABRE{} implementation,
(ii)~authoring benchmark scripts and aggregating per-circuit JSON
results into the tables of Section~\ref{sec:experiments},
(iii)~drafting the LaTeX sources for the architecture and
teleport-candidate figures
(Figures~\ref{fig:arch} and~\ref{fig:tele-candidates}), and
(iv)~copy-editing prose throughout the manuscript.  All
AI-generated content was reviewed, verified, and corrected by the
author, who takes full responsibility for the correctness and
originality of the work.

%%─────────────────────────────────────────────────────────────────────────────
\bibliographystyle{IEEEtran}
\bibliography{dsabre}

\begin{IEEEbiography}[{\includegraphics[width=1in,height=1.25in,clip,keepaspectratio]{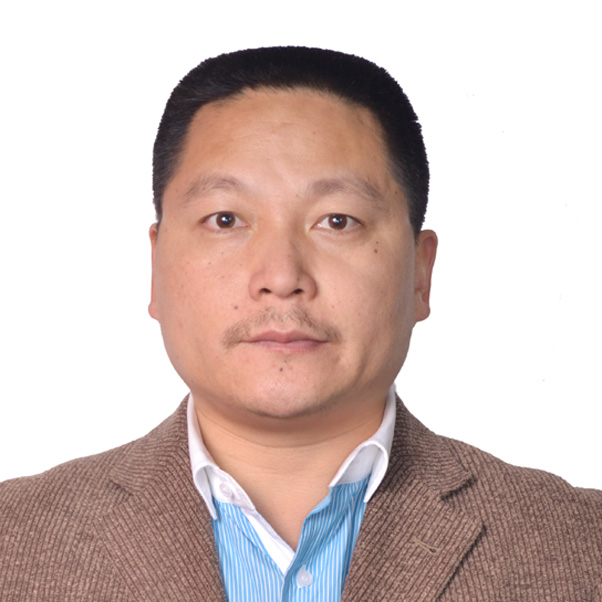}}]{Sanjiang Li}
received his B.Sc. in mathematics from Shaanxi Normal University in 1996 and his Ph.D. in mathematics from Sichuan University in 2001. He is currently a professor at the Centre for Quantum Software and Information at the University of Technology Sydney (UTS). Prior to joining UTS, he worked in the Department of Computer Science and Technology at Tsinghua University from 2001 to 2008. His primary research interests include knowledge representation, artificial intelligence, and quantum circuit compilation.
\end{IEEEbiography}

\end{document}